\documentclass[pre,twocolumn,showpacs,floatfix,amsmath,amsfonts]{revtex4}

\newcommand{\cU}{{\cal U}}
\newcommand{\cG}{{\cal G}}

\usepackage{epsfig}
\usepackage{graphicx}
\usepackage{amsmath}
\usepackage{amssymb}
\usepackage{dcolumn}
\begin{document}

\title{Generalized Neighbor-Interaction Models Induced by Nonlinear Lattices.}

\author{F. Kh. Abdullaev$^{1}$}
\email{fatkh@uzsci.net}

\author{Yu. V. Bludov$^{2}$}
\email{bludov@cii.fc.ul.pt}

\author{S. V. Dmitriev$^{3}$}
\email{dmitriev.sergey.v@gmail.com}

\author{P. G. Kevrekidis$^{4}$}
\email{kevrekid@math.umass.edu}

\author{V. V. Konotop$^{2,5}$}
\email{konotop@cii.fc.ul.pt}

\affiliation{ $^1$Instituto de F\'{\i}sica Te\'orica, UNESP, Rua Pamplona, 145, Sao Paulo, Brasil
\\
$^2$Centro de F\'{\i}sica Te\'orica e Computacional, Universidade de Lisboa, Complexo Interdisciplinar, Avenida Professor Gama Pinto 2, Lisboa
1649-003, Portugal
\\
$^3$General Physics Department, Altai State Technical University, 656038 Barnaul, Russia
\\
$^4$ Department of Mathematics and Statistics, University of Massachusetts, Amherst, MA 01003 USA
\\
$^5$Departamento de F\'{\i}sica, Universidade de Lisboa, Campo Grande, Ed. C8, Piso 6, Lisboa 1749-016, Portugal }

\begin{abstract}
It is shown that the tight-binding approximation of the nonlinear
Schr\"odinger equation with a periodic linear potential and periodic in space
nonlinearity coefficient gives rise to a number of nonlinear lattices with complex, both linear and nonlinear, neighbor interactions. The obtained
lattices present non-standard possibilities, among which we mention a
quasi-linear  regime, where the pulse dynamics obeys essentially
the linear Schr{\"o}dinger equation.
We analyze the properties of such models both in connection with their
modulational stability, as well as in regard to the existence and stability
of their localized solitary wave solutions.
\end{abstract}

\maketitle

\section{Introduction}

It is generally recognized that mapping of a nonlinear evolution problem, described by a partial differential equation,
into a simplified lattice, representing a set of coupled ordinary differential equations, appears to be a useful tool
either for numerical (or semi-analytical in the appropriate limits) study of the dynamics or for bringing intuitive
understanding of the factors dominating the behavior of the systems. Examples of such approach are well known for a
long time in solid state physics~\cite{solid} (the description of an electron in a crystal in the tight-binding
approximation), in optics~\cite{optics} (the description of the electric field in arrays of waveguides), and more
recently in the mean-field theory of Bose-Einstein condensates loaded in optical lattices (see e.g. ~\cite{BK,MO} for relevant reviews). In
all mentioned cases, the periodicity is usually associated with the linear properties of the system and the respective
dynamics is approximately described by the discrete nonlinear Schr\"odinger (DNLS) equation.

On  the other hand, there has recently been an increasing interest in
studying nonlinear models, where the nonlinearity is also
periodically modulated in space. Applications of such models extend from
the propagation of electromagnetic waves in stratified
media~\cite{Fibich} to condensates of bosons~\cite{bose} and
condensates of boson-fermion mixtures~\cite{BluKon} in optical
lattices. It turned out that the spatially dependent nonlinearity
may dramatically change properties of the system, in particular
the regions of existence of coherent localized structures and
especially their corresponding stability properties.

Natural questions that arise in this context are the mapping of
the respective evolution equation into nonlinear lattice and
the description of the existence, stability, and dynamics
properties of such lattices. In the present
paper we consider both of these issues. While the first issue is technical
and can be straightforwardly addressed by means of the Wannier
function expansion, as it was suggested in~\cite{AKKS}, the study of
the properties of the emerging lattices is a much richer
problem in the present setting; its richness stems from the fact that spatially
dependent nonlinearity gives rise to complex nonlinear
inter-site interactions. Such additional forms of nonlinearity
as the ones extracted below can
significantly change the dynamical properties of the discrete
system as it was shown in earlier
research devoted to the spin waves in magnetic
systems~\cite{spin}, electromagnetic waves in waveguide
arrays~\cite{OJE,OGJC}  and to applications to arrays of
Bose-Einstein condensates~\cite{ST}. In the above mentioned studies of
nonlinear lattices, however, one common feature was of crucial
importance -- that was the presence of dominant (or at least
significant) on-site nonlinearity, which e.g., in the case of a BEC
loaded in an optical lattice is typically about two orders of
magnitude larger than the hopping nonlinearity.  In the present
paper, we systematically derive and consider
a far more general class of lattice evolution equations,
including the cases where the on-site nonlinearity is exactly
zero.

The organization of the paper is as follows. In Sec.~\ref{model}
we deduce the relevant lattice dynamical models with inter-site
nonlinearity starting with the evolution equation of the nonlinear
Schr{\"o}dinger (NLS) type with a spatially periodic potential and
periodic nonlinearity. In Sec. \ref{Sec:MI}, we conduct the
modulational stability analysis of the derived models. In sections
\ref{Sec:linear} and \ref{Sec:nonlinear}, we examine the
respective dynamical properties of the derived quasi-linear and
nonlinear models, while Sec. \ref{Sec:conclusions} summarizes our
findings and presents our conclusions, as well as directions of
potential future interest.

\section{Model equations}
\label{model}

\subsection{One-band approximation}

We start with the one-dimensional NLS equation
\begin{eqnarray}
\label{psi_fin}
i\frac{\partial \psi}{\partial t}=-\frac{\partial^2\psi}{\partial x^2}+\cU(x)\psi+\cG(x)|\psi|^2\psi,\label{boson1D}
\end{eqnarray}
where $\cU(x)$ and  $\cG(x)$ are the coordinate-dependent
linear and nonlinear potentials respectively, both considered to be
$\pi$-periodic functions: $\cU(x)=\cU(x+\pi)$ and $\cG(x)=\cG(x+\pi)$. We
concentrate on the cases where the linear potential is an even
function $\cU(x)=\cU(-x)$, while the nonlinearity may be either
even or odd: $\cG(x)=\sigma \cG(-x)$ (hereafter $\sigma=\pm 1$).


In order to map Eq. (\ref{psi_fin}) into a lattice equation we
follow \cite{AKKS}. To this end, we introduce the linear eigenvalue
problem
\begin{eqnarray}
 -\frac{d^2\varphi_{\alpha q}(x)}{dx^2} +
        \cU(x) \varphi_{\alpha q}(x)={\mathcal E}_{\alpha q}\varphi_{\alpha q}(x)
        \label{eq:1deigen}
\end{eqnarray}
where $\varphi_{\alpha q}(x)$ is a Bloch function, $\alpha\ge 1$ and $q$ stand for the band number and for the wavenumber in the first Brillouin
zone: $q\in[-1,1]$, and define the Wannier functions
\begin{eqnarray}
w_{n\alpha}(x)=\frac{1}{\sqrt{2}}\int_{-1}^1{\varphi_{\alpha q}(x)e^{-i\pi nq}\,dq},
\end{eqnarray}
which constitute an orthonormal set of real and exponentially
decaying functions~\cite{Kohn}.

We seek the solution of Eq.
(\ref{psi_fin}) in the form of a series
\begin{eqnarray}
\label{expan1}
\psi(x,t)=\sum_{n,\alpha}{c_{n\alpha}(t)}w_{n\alpha}(x).
\end{eqnarray}
For the next consideration we notice
that the Wannier functions of $\alpha$-th band possess either even or odd parity
$w_{0\alpha}(x)=(-1)^{1+\alpha}w_{0\alpha}(-x)$ and are characterized by the property
$w_{n\alpha}(x)=w_{0\alpha}(x-n\pi)$.


Now we make the most crucial approximation of our model,
namely that the continuum Eq. (\ref{boson1D}) can be
accurately described within the one-band approximation.
As it was shown in~\cite{AKKS},
this assumption fails to describe the original continuous model
when one studies dynamical processes associated with  the
generation of the frequencies belonging to the higher bands.
However, it is reasonably accurate in describing static solutions (in
particular, localized modes) as well as their stability. Also, the
lattices of generalized neighbor interactions derived
below
within the framework of the one-band approximation, are of
interest in their own right, per their particularities and
differences in comparison to other models of similar type;
cf.~\cite{OJE,OGJC}.

We thus assume that only one band, say $\alpha$-th one, is populated.
Now, substituting the expansion (\ref{expan1}) in Eq.
(\ref{psi_fin})  we arrive at the equation (see~\cite{AKKS,BK}
for more details)
\begin{eqnarray}
i\dot{c}_{n\alpha}-
        c_{n\alpha}\omega_{0\alpha}-\left(c_{n-1,\alpha}+c_{n+1,\alpha}\right)\omega_{1\alpha} -\nonumber\\
        \sum_{n_1,n_2,n_3}c_{n_1\alpha}
        \bar{c}_{n_2\alpha}c_{n_3\alpha}W^{nn_1n_2n_3}_{\alpha\alpha\alpha\alpha}=0,
\label{eq:psi1d-aaa}
\end{eqnarray}
where
\begin{eqnarray}
&&
W^{nn_1n_2n_3}_{\alpha\alpha_1\alpha_2\alpha_3}=
\nonumber\\
&&
\int_{-\infty}^{\infty}\cG(x)w_{n\alpha}(x)w_{n_1\alpha_1}(x)w_{n_2\alpha_2}(x)w_{n_3\alpha_3}(x) dx
\end{eqnarray}
are the nonlinear overlap integrals, $\omega_{n\alpha}$ are the
coefficients of the Fourier series expansion of the eigenvalue
${\mathcal E}_{\alpha q}$:
\begin{eqnarray}
{\mathcal E}_{\alpha q}=\sum_{n}\omega_{n\alpha}e^{i\pi nq}\,, \quad \omega_{n\alpha}=\frac{1}{2}\int_{-1}^1{\mathcal E}_{\alpha q}e^{-i\pi nq}dq;
\end{eqnarray}
the overbar stands for complex conjugation, and an overdot stands for the derivative with respect to time. In Eq. (\ref{eq:psi1d-aaa}) we have
taken into account that in a general situation, for a periodic potential $\cU(x)$ of rather large amplitude, the lowest bands are very narrow, and
hence the Fourier coefficients $\omega_{n\alpha}$ decay rapidly with
increasing  $n$ (so that $|\omega_{0\alpha}| \gg |\omega_{1\alpha}| \gg
|\omega_{2\alpha}|$), which, in turn, allows us to neglect the coefficients $\omega_{n\alpha}$ with $n\ge 2$.

For the lowest bands the Wannier functions are well localized on the scale of one lattice period (and can be reasonably
well approximated by the eigenstates of the linear oscillator) and thus the overlap integrals involving next-nearest
neighbors (i.e., lattice minima separated by two lattice maxima) are negligibly small. This allows us to drop also the
terms involving $W_{\alpha_1\alpha_2\alpha_3\alpha_4}^{n_1n_2n_3n_4}$ with at least one pair of the upper indices
satisfying $|n_j-n_k|\geq 2$. We however emphasize, that it is of crucial importance to leave the nonlinear terms with
hopping between the neighbor sites, which for specific choices of the nonlinear interactions $\cG(x)$ can be comparable
with or even stronger than the on-site nonlinearity (see below).

Now we use the symmetry of the integrals $W$ with respect to
permutations of the indices and introduce
\begin{eqnarray}
    \label{prop_W}
    \begin{array}{l}
    W_0=W_{\alpha\alpha\alpha\alpha}^{nnnn}=W_{\alpha\alpha\alpha\alpha}^{0000},
    \\     W_1=W_{\alpha\alpha\alpha\alpha}^{n,n-1,n-1,n-1}=\sigma W_{\alpha\alpha\alpha\alpha}^{n,n,n,n-1}=W_{\alpha\alpha\alpha\alpha}^{1000} ,
    \\ W_2= W_{\alpha\alpha\alpha\alpha}^{n,n,n-1,n-1}=W_{\alpha\alpha\alpha\alpha}^{1100},
    \end{array}
\end{eqnarray}
where $\sigma=1$ and $\sigma=-1$ for $\cG(x)$ even and odd and
\begin{eqnarray}
\label{W_def} W_j=\int_{-\infty}^{\infty}\cG(x)w_{1\alpha}^j(x)w_{0\alpha}^{4-j}(x)dx \, \quad j=0,1,2.
\end{eqnarray}
In the case of odd nonlinearity ($\sigma=-1$) the terms $W_0$ and $W_2$ are always equal to zero due to the fact that the integrand in (\ref{W_def})
is odd with respect to the points $X=0$ and $X=\pi/2$, correspondingly. We thus arrive at the equation
\begin{eqnarray}
i{\dot c}_n= \omega_{0}c_n+\omega_{1}(c_{n-1}+c_{n+1})  +W_0|c_{n}|^2c_{n}
\nonumber\\
+ W_1\left(|c_{n-1}|^2c_{n-1}+\sigma\overline{c}_{n-1}c_{n}^2+2\sigma|c_{n}|^2c_{n-1}
\right.\nonumber\\
\left. + 2|c_{n}|^2c_{n+1}+\overline{c}_{n+1}c_{n}^2+\sigma |c_{n+1}|^2c_{n+1}\right)
\nonumber \\
+W_2\left(2|c_{n-1}|^2c_{n}+\overline{c}_{n}c_{n-1}^2+\overline{c}_{n}c_{n+1}^2+ 2|c_{n+1}|^2c_{n}\right),
\label{eq:psi1d-exp1}
\end{eqnarray}
where we have dropped the zone index $\alpha$ (e.g. $c_{n,\alpha}$ is redefined as $c_n$, etc.)

Eq. (\ref{eq:psi1d-exp1}) is the main discrete model studied in
the present paper. We notice that it has a Hamiltonian structure:
$i\dot{c}_n=\partial H/\partial \bar{c}_n$ with the Hamiltonian
\begin{eqnarray}
    \label{Ham}
    H=\sum_n\left[\omega_{0}|c_n|^2+\omega_{1}\left(c_{n-1}\bar{c}_n+\bar{c}_{n-1}c_n\right)
    +\frac{W_0}{2}|c_n|^4
        \right.
    \nonumber \\
    +
    W_1\left(|c_{n-1}|^2+\sigma |c_{n}|^2\right)\left(c_n\bar{c}_{n-1}+\bar{c}_nc_{n-1}\right)
    \nonumber \\
    \left. +   2W_2|c_{n-1}|^2|c_{n}|^2+\frac{W_2}{2}\left(c_n^2\bar{c}_{n-1}^2+\bar{c}_n^2c_{n-1}^2\right)
    \right]
\end{eqnarray}
and with the standard Poisson brackets. Another integral of motion
is the sum $N=\sum_n|c_n|^2$, reflecting the conservation of the
``number of atoms'' (in keeping with the BEC motivation of our analysis)
of the original Eq. (\ref{boson1D}).

\subsection{Particular cases}
\label{sec:part_case}

The deduced model  (\ref{eq:psi1d-exp1}) allows for a number of
interesting particular cases. First of all, however we notice that in the
case of odd nonlinearity function, i.e., for $\sigma=-1$, neither purely
even ($c_n=c_{-n}$), nor purely odd ($c_n=-c_{-n}$) solutions can exist.
 This follows directly from the symmetry of Eq. (\ref{eq:psi1d-exp1}) [or
Eq.~(\ref{psi_fin})].

For illustration of the localized solutions we use the potential $\cU(X)=-3\cos(2X)$ (both for the detailed calculations of this section,
and  for the numerical simulations of the following sections).
Also we restrict our considerations to the first band, i.e. we take $\alpha=1$. In that case, the linear overlap coefficients for the first band
are computed as $\omega_{0}\approx -0.839$, $\omega_{1}\approx -0.051$, and $\omega_{2}\approx 0.002$~\footnote{The numerical values of the Fourier
coefficients, of the Wannier functions and of the overlap integrals,
presented in this paper were obtained by
the software developed by G. L. Alfimov.}.

One can distinguish five cases as follows:

\smallskip
{\em Case 1}: $W_0=W_1=W_2=0$. This is a quasi-linear case, which
is made possible, for example by an odd nonlinearity ($\sigma=-1$) of the form
\begin{eqnarray}
\label{linear} \cG(x)=\sin(2x)-1.3706\sin(4x)
\end{eqnarray}
(obviously a number of possible realizations of this and other cases reported below is naturally unlimited).  Our approach both in this example and
below is motivated by the nature of the lattice that we wish to construct (i.e., by the type of overlap integral that we wish to preserve or
eliminate). For instance, in this example, the odd nonlinearity guarantees that $W_0=W_2=0$, while the expression of Eq. (\ref{linear}) uses one
tunable parameter (the amplitude of the second harmonic) to achieve $W_1=0$.

\smallskip

 {\em Case 2}: $W_0=W_2=0$, $W_1\neq 0$, is achieved e.g. by choosing
 \begin{eqnarray}
 \label{case2_odd}
 \cG(x)=10\sin(2x)
 \end{eqnarray}
($\sigma=-1$). Now the lattice model (\ref{eq:psi1d-exp1}) is reduced to
\begin{eqnarray}
    \label{case2_even}
&&  i{\dot c}_n = \omega_{0}c_n+\omega_{1}(c_{n-1}+c_{n+1})
\nonumber\\
&&+ W_1\left(|c_{n-1}|^2c_{n-1}-\overline{c}_{n-1}c_{n}^2-2|c_{n}|^2c_{n-1}
\right.
\nonumber\\
&&+ \left.    2|c_{n}|^2c_{n+1}+\overline{c}_{n+1}c_{n}^2- |c_{n+1}|^2c_{n+1}\right),
\end{eqnarray}
and in the case at hand $W_1\approx 0.045$.

Alternatively, this case can be realized by choosing even nonlinearity ($\sigma=1$)
\begin{eqnarray}
\cG(x)=0.0275+4.809\cos(2x)-10\cos(4x).
\end{eqnarray}
Now $W_1\approx0.012$ and the lattice equation reads:
\begin{eqnarray}
    \label{case2}
&&  i{\dot c}_n = \omega_{0}c_n+\omega_{1}(c_{n-1}+c_{n+1})
\nonumber\\
&&+ W_1\left(|c_{n-1}|^2c_{n-1}+\overline{c}_{n-1}c_{n}^2+2|c_{n}|^2c_{n-1} \right.
\nonumber\\
&&+ \left.    2|c_{n}|^2c_{n+1}+\overline{c}_{n+1}c_{n}^2+ |c_{n+1}|^2c_{n+1}\right).
\end{eqnarray}

\smallskip

{\em Case 3}: $W_0=W_1 =0$, $W_2\neq 0$ is obtained, for instance, for the even nonlinearity ($\sigma=1$)
\begin{eqnarray}
\cG(x)=-23.836+48.882\cos(2x)-37.778\cos(4x),
\end{eqnarray}
for which $W_2\approx -0.0136$. The model (\ref{eq:psi1d-exp1}) is now simplified
\begin{eqnarray}
i{\dot c}_n=\omega_{0}c_n+\omega_{1}(c_{n-1}+c_{n+1})
\nonumber\\
+W_2\left(2|c_{n-1}|^2c_{n}+\overline{c}_{n}c_{n-1}^2+\overline{c}_{n}c_{n+1}^2+ 2|c_{n+1}|^2c_{n}\right).
\label{case3}
\end{eqnarray}

\smallskip

{\em Case 4}: $W_0=0$, $W_{1,2}\neq 0$ can be achieved by using an even nonlinearity ($\sigma=1$)
\begin{eqnarray}
\cG(x)=7.795-10\cos(2x).
\end{eqnarray}
Now the overlap integrals are as follows $W_1\approx 0.0148$ and $W_2\approx0.0045$. The lattice model then reads
\begin{eqnarray}
i{\dot c}_n= \omega_{0}c_n+\omega_{1}(c_{n-1}+c_{n+1})
\nonumber\\
+ W_1\left(|c_{n-1}|^2c_{n-1}+\overline{c}_{n-1}c_{n}^2+2|c_{n}|^2c_{n-1}
\right.\nonumber\\
\left. + 2|c_{n}|^2c_{n+1}+\overline{c}_{n+1}c_{n}^2+ |c_{n+1}|^2c_{n+1}\right)
\nonumber \\
+W_2\left(2|c_{n-1}|^2c_{n}+\overline{c}_{n}c_{n-1}^2+\overline{c}_{n}c_{n+1}^2+ 2|c_{n+1}|^2c_{n}\right).
\label{case4}
\end{eqnarray}

\smallskip

{\em Case 5}: $|W_1|,\,|W_2| \ll |W_0|$ is the standard case of
the on-site nonlinearity (the DNLS equation):
\begin{eqnarray} \label{case5}
   i{\dot c}_n= \omega_{0}c_n+\omega_{1}(c_{n-1}+c_{n+1})
   +W_0|c_{n}|^2c_{n}.
\end{eqnarray}
This form of the lattice dynamical model is obtained for
generic nonlinearities, Eq.(\ref{case5}) is well studied in the
literature (see e.g. \cite{DNLS_review}) and that is why
it will not be addressed in this paper.

\section{Modulational Instability}
\label{Sec:MI}

As it is customary we start with the analysis of the modulational instability of plane-wave solutions of Eq. (\ref{eq:psi1d-exp1}) (for a general
study of the modulational instability of the plane wave background in the DNLS-type equations see e.g.~\cite{KivsharSalerno,KT,ADG}). This stability
analysis is performed not only because it is of interest in its own right but also because the solitary wave solutions that we plan on constructing
for the above presented lattices should be produced on a dynamically stable background. Using the plane wave solution of the form:
\begin{equation}
 c_n = Fe^{i(qn - \omega t)},\label{eq:mi-anz}
\end{equation}
where $F$ is a constant amplitude, we
obtain the dispersion relation
(in the absence of the previously considered cubic onsite terms)
\begin{eqnarray}
\label{omega}
\omega &=& \omega_{0} + 2\omega_{1}\cos(q) + 4W_1F^2(\sigma+1)\cos(q) + \nonumber\\
&& 2W_{2}F^2[2 + \cos(2q)].
\end{eqnarray}
To examine the linear stability, we perturb the plane wave solution in the form:
\begin{eqnarray}
c_{n} = (F +Ae^{i(Qn-\Omega t)} + \bar{B}e^{-i(Qn-\Omega t)})e^{i(qn - \omega t)},
\end{eqnarray}
with $|A|\,, |B| \ll |F|$ and linearize with respect to $A$ and $B$. As a result
we obtain two branches of the linear excitations $\Omega_{1,2}(Q)$:
\begin{eqnarray}
\label{disp_rel}
\Omega_{1,2} = M_- \pm \sqrt{(M_+ - D)^2
- F^4\Delta},
\end{eqnarray}
where
\begin{eqnarray*}
D &=& \omega-\omega_0-4W_1(1+\sigma)F^2\cos(q)- \nonumber\\
&& 4W_2 F^2\left(\cos(Q)+1\right)\,,
\\
M_- &=& -2\left(\omega_{1}+2W_1(1+\sigma)F^2\right)\sin(Q)\sin(q) -\nonumber\\
&& 4W_2 F^2 \sin(Q)\sin(2q),
\\
M_+ &=& 2\left(\omega_{1}+2W_1(1+\sigma)F^2\right)\cos(Q)\cos(q) + \nonumber\\
&& 4W_2 F^2 \cos(Q)\cos(2q)\,,
\\
\Delta &=& 4\left[W_1(1+\sigma)(1+\cos(Q))\cos(q)+ \right.\nonumber\\
&& \left.W_2 (2\cos(Q) + \cos(2q))\right]^2 + \nonumber\\
&& 4(1-\sigma)^2\sin^{2}(q)W_1^2 (1-\cos(Q))^2.
\end{eqnarray*}
Let us consider different special cases for $\sigma = -1$ (the case $\sigma=1$ is investigated in \cite{OJE}).

\paragraph{Homogeneous background} is described by
$q=0$. Now the two branches of the solutions collapse and the dispersion relation acquires the form
\begin{eqnarray}
\label{q0} \Omega^2 &=& 16(\omega_{1}+2W_2 F^2)\sin^2\left(\frac{Q}{2}\right)
 \nonumber \\
&\times& \left[(\omega_{1}+ 6W_2 F^2)\sin^{2}\left(\frac{Q}{2}\right) -3W_2 F^2\right].
\end{eqnarray}

The carrier field is stable if and only if the two following conditions are satisfied
\begin{eqnarray} \label{NontrivStability}
   \left(\omega_{1}+2W_2F^2\right)W_2 \le 0, \nonumber \\
   \left(\omega_{1}+2W_2F^2\right)\left(\omega_{1}+3W_2F^2\right) \ge 0,
\end{eqnarray}
where the first of these conditions demands non-negativity of the coefficient $k$ in the expansion of Eq. (\ref{q0}) of
the form of $\Omega^2(Q)=kQ^2$, which is valid for small $Q$. The second condition demands non-negativity of
$\Omega^2(Q)$ at the zone boundary, $Q=\pi$.

We point out that the long wavelength excitations'
group velocity dispersion is given by
\begin{equation}
    \frac{\partial^2 \Omega}{\partial Q^2}\Bigg|_{Q=0}=2W_2\left[2W_2F^4+\left(8W_2-6+\omega_{1}\right)F^2+4\omega_{1}\right]
\end{equation}
and takes zero values for
$$F=\left\{\frac{6-\omega_1-8W_2}{4W_2}\pm\sqrt{\left(\frac{6-\omega_1-8W_2}{4W_2}\right)^2-\frac{2\omega_1}{W_2}}\right\}^{1/2}.$$ This last condition
determines the domain of the parameters where shock waves can be observed~\cite{shock}.

\paragraph{Staggered background} corresponds  to $q=\pi$. The dispersion relation is
\begin{eqnarray}\label{qpi}
\Omega^2 &=& 16 (\omega_{1} - 2W_2 F^2)\sin^2\left(\frac{Q}{2}\right)\nonumber \\
&\times& \left[(\omega_{1}- 6W_2 F^2)\sin^{2}\left(\frac{Q}{2}\right)
+3W_2 F^2\right].
\end{eqnarray}
Similar to (\ref{NontrivStability}) we introduce stability criteria in the form
\begin{eqnarray} \label{NontrivStability-pi}
   \left(\omega_{1}-2W_2F^2\right)W_2 \ge 0, \nonumber \\
   \left(\omega_{1}-2W_2F^2\right)\left(\omega_{1}-3W_2F^2\right) \ge 0.
\end{eqnarray}

Note that the dispersion relations for $q=0,\pi$  do not depend on
the coefficient $W_1$.

\paragraph{Phase alternating background} where $q = \pi/2$. The dispersion relation is
\begin{eqnarray}
\Omega_{1,2} = -2\omega_{1}\sin(Q) \pm \nonumber\\
4\sqrt{2}F^2\sin\left(\frac{Q}{2}\right)\sqrt{ \left(W_2^2+W_1^2\right)\cos(Q)-W_1^2}.
\end{eqnarray}
The condition for the presence of modulational instability in this case is
\begin{equation}
\cos(Q) < \frac{W_1^2}{W_1^2+W_2^2}.
\end{equation}

\section{Quasi-linear Model (Case 1)}
\label{Sec:linear}

Let us now turn to the quasi-linear model of Case 1. We notice that the term
``quasi-linear'' is used in order to emphasize the existence of solely
higher order
contributions from more distant neighbors which are small but not
necessarily zero. This provides us with an excellent benchmark
of our derivation
since in the discrete linear case, the dynamical equation can be solved explicitly and subsequently compared to the full results of the original
partial differential equation (for which the discrete model was developed as an approximation). In particular, it is well-known that the linear
discrete case, with a compactum of initial data $c_{n}(0)=A \delta_{n0}$ has a solution of the form
\begin{eqnarray}
c_n(t)=A \left(-i\right)^n \exp(-i\omega_{0} t)
    J_n(2 \omega_{1} t)
\label{linear1}
\end{eqnarray}
where $J_n$ is the Bessel function of order $n$.

\begin{figure}[htbp]
\includegraphics[width=5cm,height=6cm,angle=0,clip]{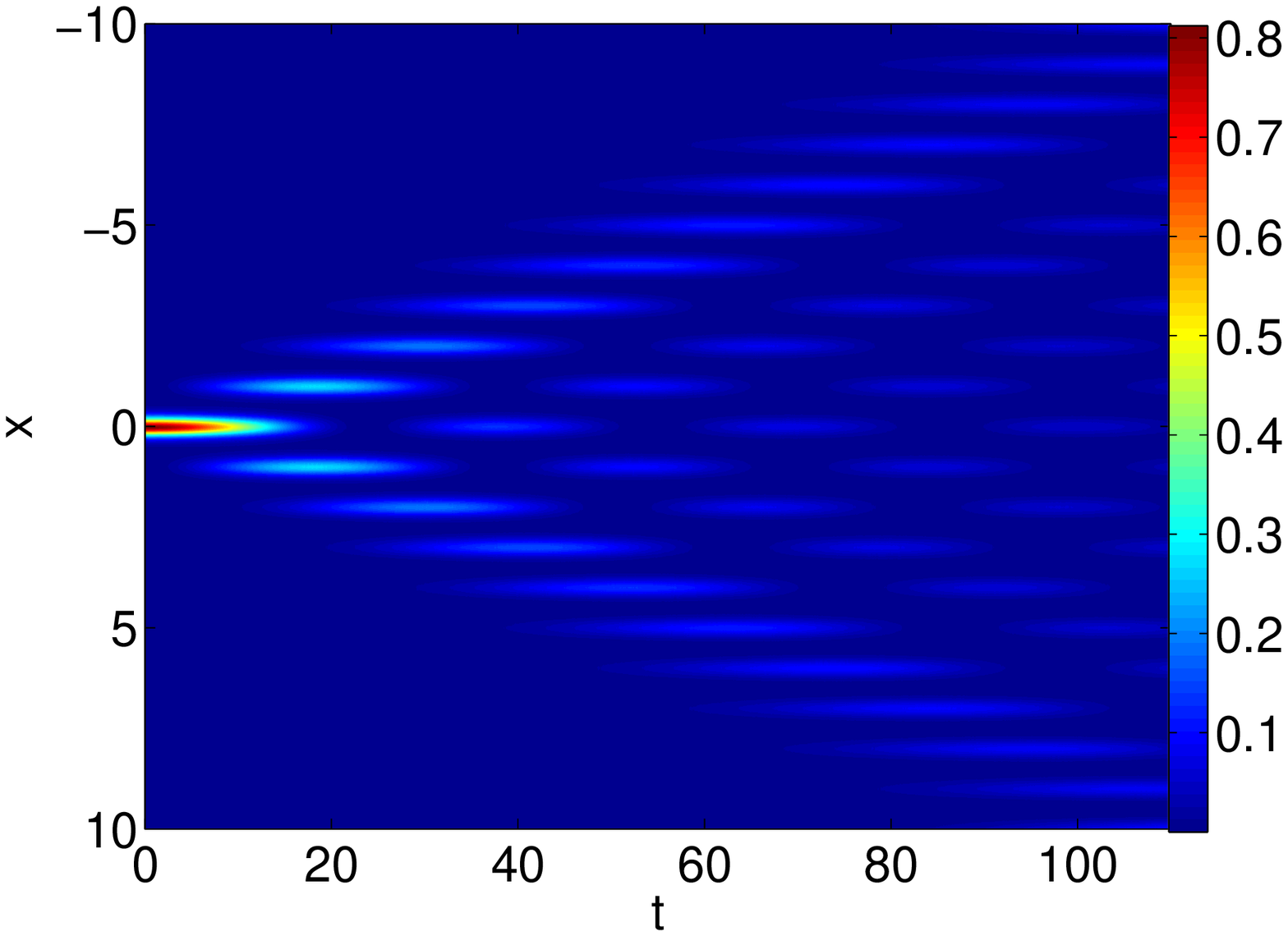}
\includegraphics[width=5cm,height=6cm,angle=0,clip]{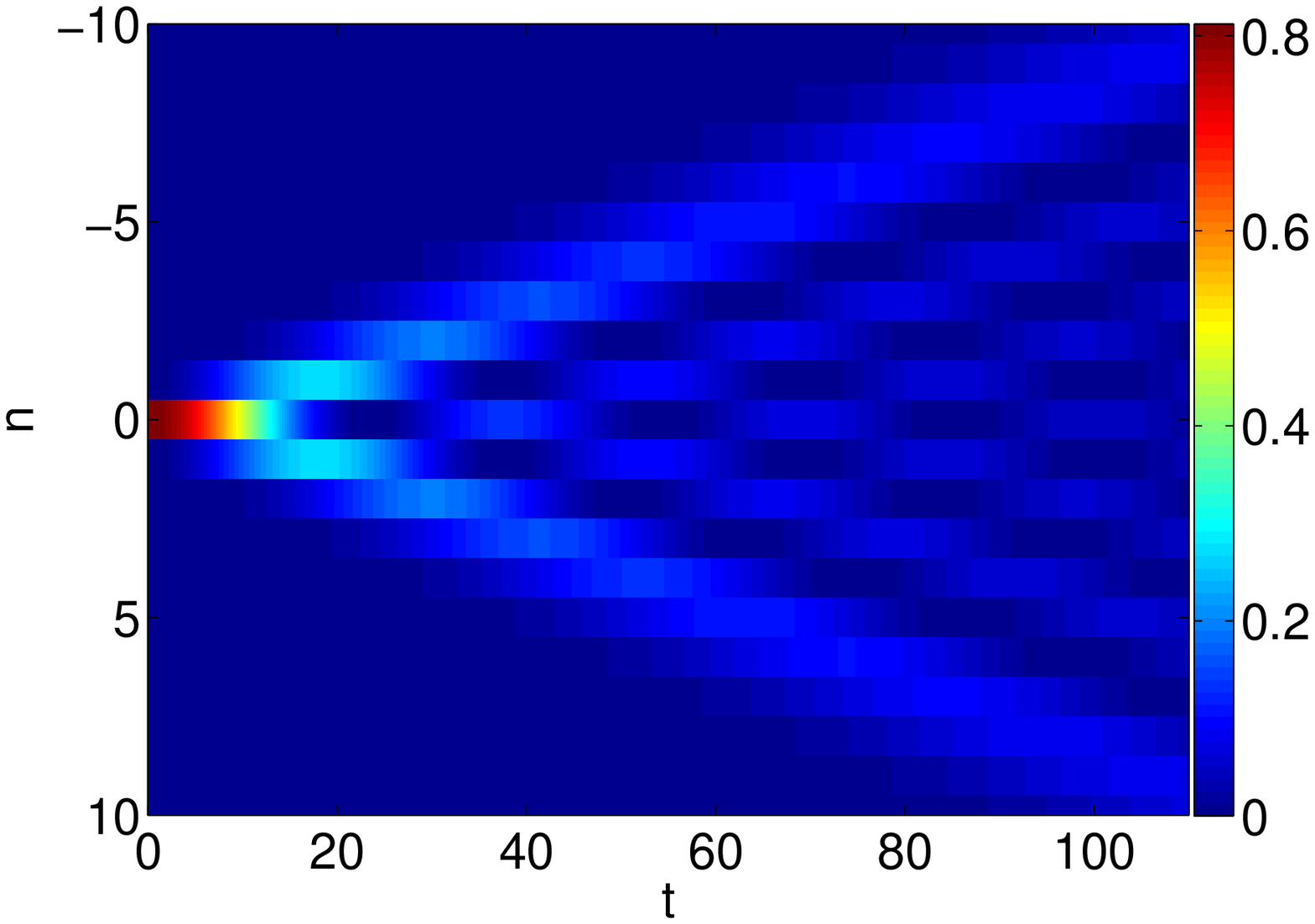}
\includegraphics[width=5cm,height=6cm,angle=0,clip]{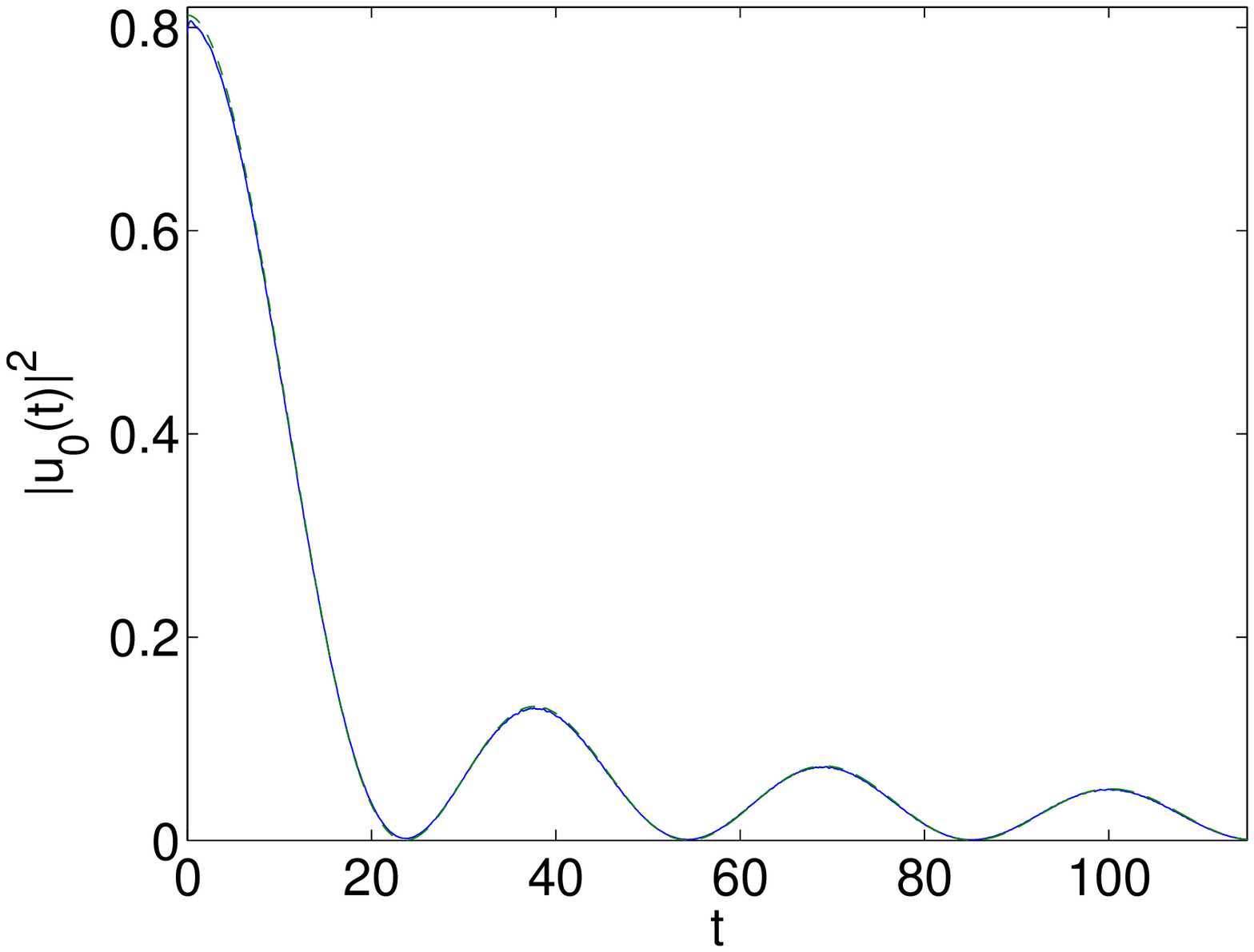}
\caption{The top and middle panel show the space-time contour
plot  of the solution
in the PDE of Eq. (\ref{boson1D}) [top panel] and the discrete model
analytical prediction (\ref{linear1}) [middle panel].
In the latter case,
space is normalized over the period of the linear potential,
so that it can be compared with the lattice results.
To accentuate the excellent agreement between analytical and
numerical results, the bottom panel shows
the time evolution of the amplitude at the
central site of the configuration compared between the
PDE numerical result (solid line) and the discrete equation
analytical result of (\ref{linear1}) (dashed line).
The two are practically {\it indistinguishable}.} \label{bfig1}
\end{figure}

We have tested this analytical prediction of the {\it discrete model}
in the partial differential equation (\ref{boson1D}) with the
``nonlinear potential'' of Eq. (\ref{linear}). The results of our numerical
simulations can be found in Fig. \ref{bfig1}, which highlights the
excellent agreement between the analytical and the numerical
results. This also serves to showcase the accuracy of the
reduction via the tight-binding approximation of the original
partial differential equation. We will hereafter focus more
on the details of these discrete models and of their solitary
wave solutions.

The remarkable accuracy of the tight-binding model in the case at hand can be easily understood. Indeed, we are dealing with a perfect lattice
(i.e. having no defects). Inter-band transitions, which are the
cause of the failure of the one-band approximation when they exist,
are only due to the
nonlinear coupling of bands, and exactly this factor is anomalously
small for the chosen nonlinearity. The first indication on this fact is
given by
the zero contributions of $W_{0,1,2}$.  Next, due to the
symmetry of the Wannier functions, and the symmetry of the
nonlinearity, one concludes that
$W_{\alpha\alpha\beta\beta}^{nnnn}=0$ for all $\alpha$ and $\beta$,
i.e., there exists no tunneling between the same sites of two
different bands. Even
more generally (also due to the symmetry),
$W_{\alpha,\beta,\beta,\alpha+2\gamma}^{n,n-m,n+m,n}=0$
(for arbitrary integers $\beta$, $\gamma$, $n$ and
$m$). For completeness we have checked the numerical values
of the other inter-band overlap integrals for the three lowest bands. The
integrals greater than $0.01$ are as follows (the ones obtained by the
symmetry reductions are not shown) $W_{2111}^{0000}=0.0465$,
$W_{2221}^{0000}=-0.1360$, $W_{2211}^{1000}=-0.0228$, $W_{3111}^{1000}=-0.0432$, $W_{3311}^{1000}=-0.0301$, $W_{1333}^{1000}=-0.0117$. The coefficients describing energy
transfer from the first to the second and the third bands are
$W_{2111}^{0000}=0.0465$, $W_{3111}^{0000}=0$, $W_{2111}^{1000}=0.0077$, $W_{3111}^{1000}=-0.0432$, $W_{2111}^{1100}=0.0004$,
$W_{3111}^{1100}=0.0011$, i.e., either have relatively small value or are
identically zero. This explains the high accuracy of the one-band
approximation.

Finally we notice, that in the case at hand the dispersion
relation (\ref{disp_rel}) takes the form independent of the wave
amplitude $F$
\begin{eqnarray}
\label{eq:qlc}
    \Omega=\pm 2\omega_1\left[\cos\left(q\right)-\cos\left(Q\mp q\right)\right],
\end{eqnarray}
which is the dispersion relation for linear phonons,
and hence no instabilities can occur.

\begin{figure}
\includegraphics{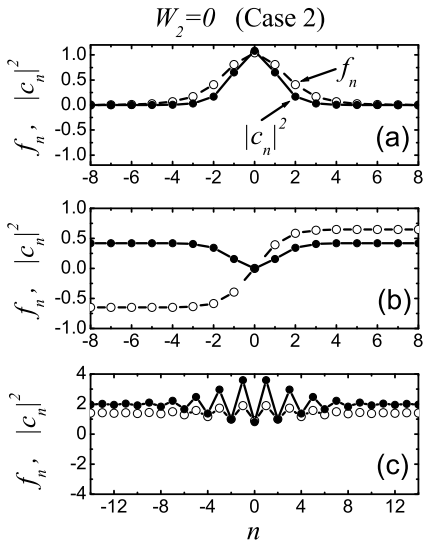}
\caption{Equilibrium soliton solutions found in the Case 2
$(W_2=0)$: (a) pulse, (b) kink, and (c) anti-dark structure with
oscillatory wave structure around the background, hereafter termed
wave. On-site structures
are presented here, but inter-site ones can also be constructed.
Solutions in (a) and (b) can be stable but we were unable to find
a stable structure of the form presented in (c) (see Sec.
\ref{Sec:stability}). Model parameters corresponding to panels (a)
to (c) are depicted by dots in Fig. \ref{Figure3} marked by the
capital letters A to C, respectively. Parameters are: (a)
$W_1=-0.0112$, $\omega=-0.99$, (b) $W_1=0.0112$, $\omega=-0.9$,
(c) $W_1=0.0112$, $\omega=-0.75$.} \label{Figure2}
\end{figure}

\begin{figure}
\includegraphics{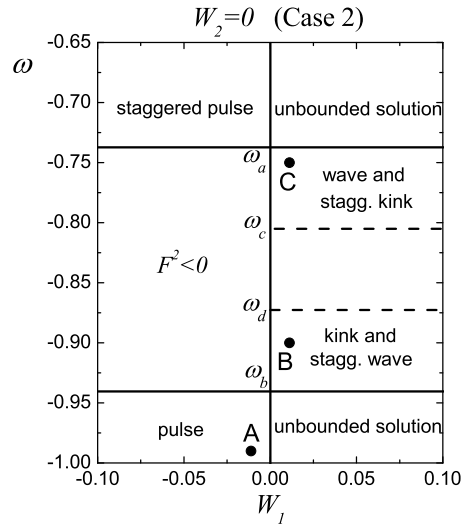}
\caption{Parameter plane $(W_1, \omega)$ in the Case 2 ($W_2=0$)
with indicated regions of existence of three types of solitons
shown in Fig. \ref{Figure2} and their staggered analogues. The
pulse solution exists for $W_1<0$ and $\omega< \omega_b$, while
the staggered pulse for $W_1<0$ and $\omega> \omega_a$. The kink
and staggered wave can exist for $W_1>0$ and $\omega_b < \omega <
\omega_c$, while the staggered kink and the wave for $W_1>0$ and
$\omega_d < \omega < \omega_a$. The values of special frequencies
$\omega_a$ to $\omega_d$ are given by Eq. (\ref{PulseTailEx}) and
Eq. (\ref{TailSign2}).} \label{Figure3}
\end{figure}

\section{Nonlinear Models (Cases 2 and 3)}   \label{Sec:nonlinear}

The above confirmation of the high accuracy of the tight-binding
model for the description of the pulse dynamics motivated us to
study the coherent structure solutions of Eq.
(\ref{eq:psi1d-exp1}) and their properties for the Cases 2 and 3
singled out in Sec.~\ref{sec:part_case}. It is convenient to join
the equations of the Cases 2 and 3 and to consider for $\sigma=1$
the model
\begin{eqnarray}
    \label{case23}
  i{\dot c}_n =
\omega_{0}c_n+\omega_{1}(c_{n-1}+c_{n+1})
\nonumber\\
+ W_1\left(|c_{n-1}|^2c_{n-1}+\overline{c}_{n-1}c_{n}^2+2|c_{n}|^2c_{n-1} \right.
\nonumber\\
+ \left.    2|c_{n}|^2c_{n+1}+\overline{c}_{n+1}c_{n}^2+|c_{n+1}|^2c_{n+1}\right)
\nonumber\\
+W_2\left(2|c_{n-1}|^2c_{n}+\overline{c}_{n}c_{n-1}^2+\overline{c}_{n}c_{n+1}^2+ 2|c_{n+1}|^2c_{n}\right). \label{case23-1}
\end{eqnarray}

Let us seek stationary solutions of Eq. (\ref{case23}) of
the form
\begin{eqnarray} \label{StationarySolution}
   c_n \left( t \right) = f_n e^{-i\omega t},
\end{eqnarray}
with real $f_n$. Using this ansatz, we obtain the equation for the
amplitudes
\begin{eqnarray} \label{SSA}
  \left( \omega_{0} - \omega \right)f_n
  + \omega_{1} \left( f_{n - 1} + f_{n + 1} \right) \\ \nonumber
  + W_1 \left[ f_{n - 1}^3  + f_{n + 1}^3  + 3f_n^2
  \left( f_{n + 1}  + f_{n - 1} \right) \right] \\ \nonumber
  + 3W_2 f_n \left( f_{n - 1}^2  + f_{n + 1}^2 \right) = 0.
\end{eqnarray}

We attempt to find the pulse (bright soliton) and kink (dark
soliton) solutions. Our strategy in searching for these solutions
will be as follows. We will first formulate the necessary
conditions for the existence of the soliton solutions considering
their carrier constant-amplitude field and also the soliton tail
solutions. This will narrow the domain of parameters where such
solutions can be expected. Then, we will attempt to construct the
desired soliton solutions and subsequently study their stability.

The staggered and non-staggered stationary solutions are connected by the following symmetry relation~\cite{ABK}: if
$f_n$ is a solution of Eq.~(\ref{SSA}) for definite $W_1$, $W_2$, and $\omega<\omega_0$ $(\omega>\omega_0)$, then
$(-1)^n f_n$ is a solution for $\widetilde{W_1}=W_1$, $\widetilde{W_2}=-W_2$, and
$\widetilde{\omega}=2\omega_0-\omega>\omega_0$ $(\widetilde{\omega}<\omega_0)$. We also note that the stability
analysis of any stationary solution can also be done, without loss
of generality,
for only, say, non-staggered carrier field. This is so because the
dynamics in the vicinity of the stationary solution is governed by Eq. (\ref{Linearized}) (see below) which is
invariant with respect to the following transformation: $\epsilon_n \rightarrow (-1)^n \epsilon_n$, $f_n \rightarrow
(-1)^n f_n$, $\omega \rightarrow 2\omega_0 - \omega$, $W_2 \rightarrow -W_2$, and $t \rightarrow -t$. Bearing this in
mind, in the following we will discuss only stationary solutions with a
non-staggered background.

\begin{figure}
\includegraphics{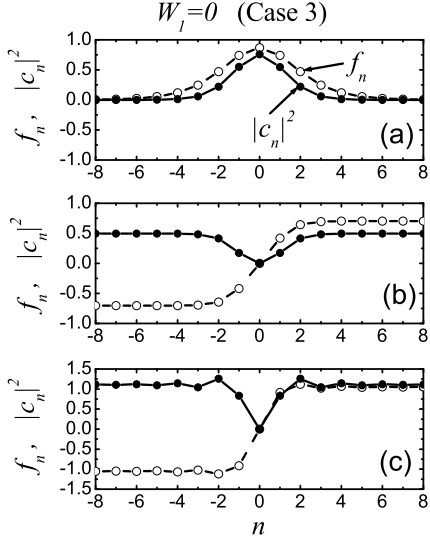}
\caption{Equilibrium soliton solutions obtained in Case 3 with
$W_1=0$: (a) pulse, (b) kink, and (c) kink with oscillatory tail,
called oscillatory kink. On-site structures are presented here,
but inter-site ones can also be constructed. All these solutions
can be stable, as it will be shown in Sec. \ref{Sec:stability}.
Model parameters corresponding to panels (a) to (c) are depicted
by dots in Fig. \ref{Figure5} marked by the capital Latin letters
A to C, respectively. Parameters are: (a) $W_2=-0.0136$,
$\omega=-0.9704$, (b) $W_2=0.0136$, $\omega=-0.9$, (c)
$W_2=0.0136$, $\omega=-0.85$.} \label{Figure4}
\end{figure}

\begin{figure}
\includegraphics{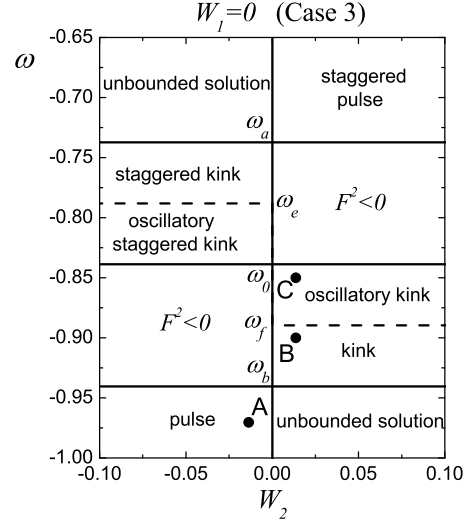}
\caption{Parameter plane $(W_2, \omega)$ in Case 3 ($W_1=0$) with
indicated regions of existence of three types of solitons shown in
Fig. \ref{Figure4} and their staggered analogues. The pulse exists
for $W_2<0$ and $\omega< \omega_b$, while the staggered pulse for
$W_2>0$ and $\omega> \omega_a$. The kink exists for $W_2>0$ and
$\omega_b < \omega < \omega_f$, while the staggered kink for
$W_2<0$ and $\omega_e < \omega < \omega_a$. Finally, the
oscillatory kink exists for $W_2>0$ and $\omega_f < \omega <
\omega_0$, while the oscillatory staggered kink for $W_2<0$ and
$\omega_0 < \omega < \omega_e$. Values of special frequencies are
given by Eq. (\ref{PulseTailEx}) and Eq. (\ref{TailSign3}).}
\label{Figure5}
\end{figure}

\subsection{ Constant amplitude solution and soliton tails }
\label{Sec:tail}

A pulse solution must satisfy the boundary conditions $|c_n|^2 \rightarrow 0$ for $n \rightarrow \pm \infty$, while for
the kink solution one must have $|c_n|^2 \rightarrow F^2 >0$ for $n \rightarrow \pm \infty$. Thus, the existence and
stability of the carrying field solution, described by formula (\ref{eq:mi-anz}) with real $F$, is a necessary
condition for the existence and stability of the soliton solutions.

One always has the trivial solution $F=0$ and from the expression
(\ref{omega}) one can have two nonzero solutions with
\begin{eqnarray} \label{NontrivF}
   F^2 = \frac{ \omega - \omega_{0} - 2\omega _{1} }{8W_1+6W_2 },
\end{eqnarray}
if the expression in the right-hand side of Eq. (\ref{NontrivF})
is positive. In the case $8W_1+6W_2 = 0$, $F$ can be arbitrary if
$\omega = \omega_{0} + 2\omega_{1}$, but we will not study this
very special case.

Substituting Eq. (\ref{NontrivF}) into Eq.
(\ref{disp_rel}) we obtain the spectrum of the carrier field with
$F^2>0$, whose stability criteria (in full analogy with Sec.
\ref{Sec:MI}) are
\begin{eqnarray} \label{NontrivStability-1}
   \left[W_1(\omega - \omega _{0} + 2\omega_{1})
   + W_2(\omega - \omega _{0} + \omega_{1}) \right]\times \nonumber\\
   \left(4W_1 + 3W_2\right)\left( \omega  - \omega _{0} - 2\omega _{1} \right) \le 0, \nonumber \\
   \left[W_1(\omega - \omega _{0} + 2\omega_{1}) + W_2(\omega - \omega _{0} + \omega_{1})
   \right]\times \nonumber \\
   \left[2W_1(\omega - \omega _{0} + 2\omega_{1}) + 3W_2(\omega - \omega _{0})
   \right] \ge 0.
\end{eqnarray}

From the asymptotic properties mentioned above, one can express
the soliton tails as
\begin{eqnarray} \label{ConstantPerturbed}
   f_n \sim F+\xi_n
\end{eqnarray}
at $|n| \to \infty$, where small $\xi_n$ are real and are independent on $t$. Substituting Eq.
(\ref{ConstantPerturbed}) into Eq. (\ref{SSA}) and linearizing with respect to $\xi_n$ one obtains
\begin{eqnarray} \label{TailEquation}
   \gamma \xi_{n - 1} + \beta \xi_n + \gamma \xi_{n + 1} = 0,
\end{eqnarray}
with
\begin{eqnarray} \label{TailCoeff}
   \beta = \omega_{0} - \omega + 12W_1 F^2 + 6W_2 F^2 , \nonumber \\
   \gamma = \omega _{1} + 6W_1 F^2 + 6W_2 F^2.
\end{eqnarray}

Seeking solutions to Eq. (\ref{TailEquation}) in the form
$$
\xi_n \sim C_{\pm}z^n, \quad n \to \pm\infty
$$
with complex $z$, we come to the characteristic equation
\begin{eqnarray} \label{CharacteristicEq}
   \gamma z^2  + \beta z + \gamma  = 0.
\end{eqnarray}
Thus, $z$ is one of the roots $z_{1,2}$
\begin{eqnarray} \label{SpecialRoots}
   z_1 = \frac{1}{z_2}= -\frac{\beta}{2\gamma}
   +\sqrt{\frac{\beta^2}{4\gamma^2} - 1}
\end{eqnarray}
providing $|\xi_n| \to 0$ as $|n|\to\infty$.

For the soliton, which is either a homoclinic or a heteroclinic of the map, generated by
Eq.~(\ref{SSA}), $\xi_n \equiv 0$ must be a hyperbolic point which
happens only if the roots $z_{1,2}$ are real, i.e., if
\begin{eqnarray} \label{TailDeterm}
   \beta^2-4\gamma^2 >0.
\end{eqnarray}

\begin{figure}
\includegraphics{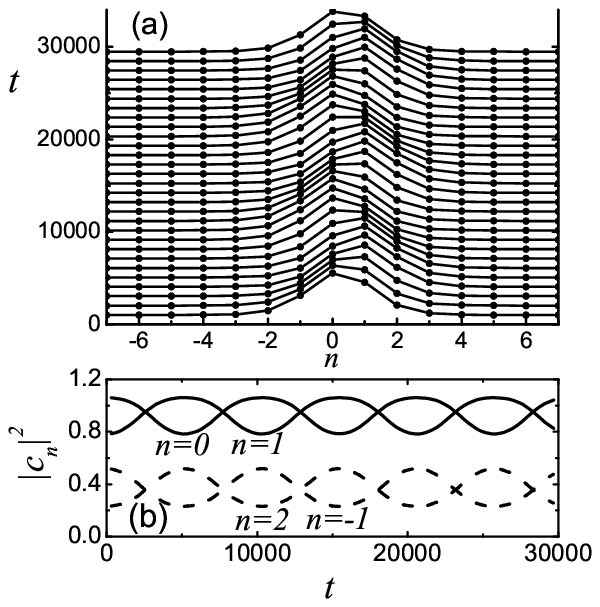}
\caption{Pulse in Case 2 ($W_2=0$). (a) Space-time evolution of
$|c_n(t)|^2$ showing the dynamics of the pulse placed at $t=0$
asymmetrically with respect to the lattice. The pulse undergoes
periodic oscillations in the vicinity of the stable inter-site
configuration. (b) Time variation of the norm of the four central
particles. Parameters: $W_1=-0.012$, $W_2=0$, $\omega= -0.99$,
which corresponds to the point A in Fig. \ref{Figure3}.}
\label{Figure6}
\end{figure}

\begin{figure}
\includegraphics{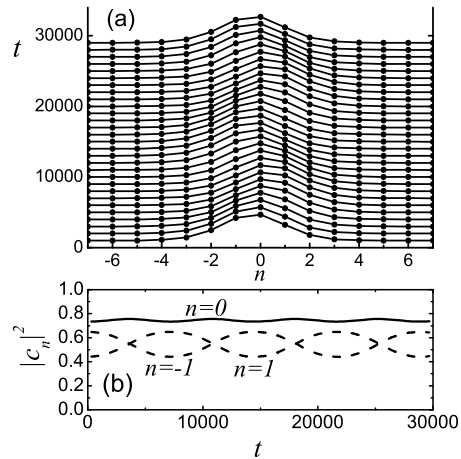}
\caption{Pulse in Case 3 ($W_1=0$). (a) Space-time evolution of
$|c_n(t)|^2$ showing the dynamics of the pulse placed at $t=0$
asymmetrically with respect to the lattice. The pulse undergoes
periodic oscillations in the vicinity of the stable on-site
configuration. (b) Time variation of the norm of the central
particle and its two nearest neighbors. Parameters: $W_1=0$,
$W_2=-0.0136$, $\omega= -0.9704$, which corresponds to the point A
in Fig. \ref{Figure5}.} \label{Figure7}
\end{figure}

The sign of $z$ specifies the type of the tail solution: $z>0$ corresponds to the tail decaying monotonically with distance from the soliton's center,
while $z<0$ means that the decaying tail solution oscillates near the carrier solution $F$.

The absolute value of $z$ characterizes the degree of localization of the
tail. If $|z|$ is small, then
the tail solution is weakly localized, otherwise it is
strongly localized. We found that, in many cases, there is a correlation between the degree of localization of the tail solution and that of the
soliton itself.

The tail solution was found from the linearized equation and it
can only provide necessary conditions for the existence of a
soliton solution. The nonlinear terms, depending on their
structure and the values of the corresponding coefficients, can
either lead to unbounded solutions or to non-localized solutions,
or to the desired bounded and localized soliton solutions.

Having a tail solution one can attempt to construct the
corresponding soliton solution. To do so, we use the tail solution
defined by Eq. (\ref{ConstantPerturbed}) to set the initial values
for $f_{n-1}$ and $f_n$ and then find $f_{n+1}$ from Eq.
(\ref{SSA}), solving the cubic (for $W_2=0$) or the quadratic (for
$W_1=0$) algebraic equation. The proper choice of the integration
constants $C_{\pm}$ systematically
allows one to obtain equilibrium on-site or
inter-site soliton solutions, if they exist \footnote{This is a
variant of the shooting method used for obtaining of localized modes
in continuous models \cite{BK,BluKon,ABK,AKS}.}.

To conclude this section we summarize the necessary conditions for
the existence of pulses and kinks.

{\em Pulse in Cases 2 and 3.} Since the carrying field for pulses
with $F=0$ always exists, there remains only one necessary
condition, namely, the condition of the existence of the tail
solution of Eq. (\ref{TailDeterm}). For $F=0$ this condition is
satisfied for any $W_1$ and $W_2$ and for both Cases 2 and 3 when
\begin{eqnarray}
\label{PulseTailEx}
\begin{array}{l}
   \omega > \omega_0 -2\omega_1 =-0.73732741 \equiv \omega_a,
   \\
   \omega < \omega_0 +2\omega_1 =-0.94041721 \equiv \omega_b.
   \end{array}
\end{eqnarray}
These conditions simply state that the localized pulses must be
located outside the phonon band of the spectrum given by the interval $[\omega_b,\omega_a]$.

The necessary conditions of the existence of a kink include the
condition of the existence of the carrier field with $F^2>0$, Eq.
(\ref{NontrivF}), the stability condition for the carrier field,
Eq. (\ref{NontrivStability-1}), and condition Eq.
(\ref{TailDeterm}) of the existence of the tail solution.

{\em Kink in Case 2.} All three necessary conditions are satisfied
when $W_1>0$ and $\omega_b <\omega< \omega_a$, and they are not
satisfied simultaneously for $W_1<0$.

{\em Kink in Case 3.} All three necessary conditions are satisfied
when $W_2>0$ and $\omega_b <\omega< \omega_0$, while for $W_2<0$
they are satisfied for $\omega_0 <\omega< \omega_a$.

We also specify the frequencies at which $z$ changes sign. In the
Case 2 the frequencies are
\begin{eqnarray} \label{TailSign2}
   \begin{array}{l}
   \omega_0 - (2/3)\omega_1 =-0.80502401 \equiv \omega_c,
   \\
   \omega_0 + (2/3)\omega_1 =-0.87272061 \equiv \omega_d,
   \end{array}
\end{eqnarray}
while in the Case 3 they are
\begin{eqnarray} \label{TailSign3}
\begin{array}{l}
   \omega_0 -\omega_1 = -0.78809986 \equiv \omega_e,
   \\
   \omega_0 +\omega_1 = -0.88964476 \equiv \omega_f.
\end{array}
\end{eqnarray}

\begin{figure}
\includegraphics{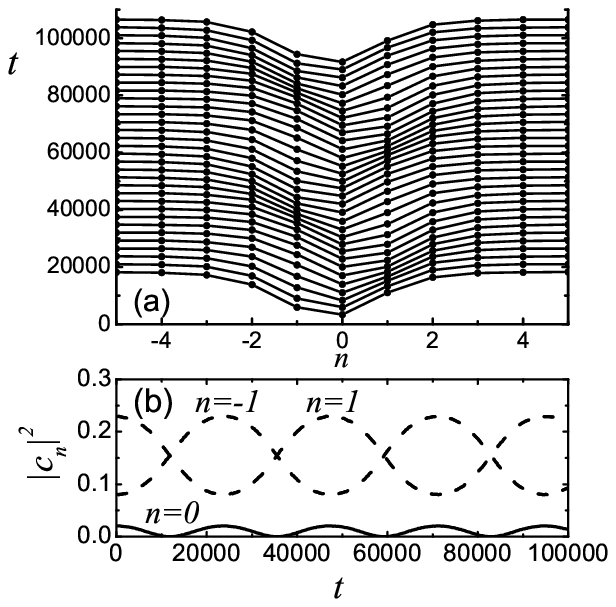}
\caption{Kink in Case 2 ($W_2=0$). (a) Space-time evolution of
$|c_n(t)|^2$ showing the dynamics of the kink placed at $t=0$
asymmetrically with respect to the lattice. The kink undergoes
periodic oscillations in the vicinity of the stable on-site
configuration. (b) Time variation of the norm of the central
particle and its two nearest neighbors. Parameters: $W_1=0.012$,
$W_2=0$, $\omega=-0.9$, which corresponds to the point B in Fig.
\ref{Figure3}. } \label{Figure8}
\end{figure}

\begin{figure}
\includegraphics{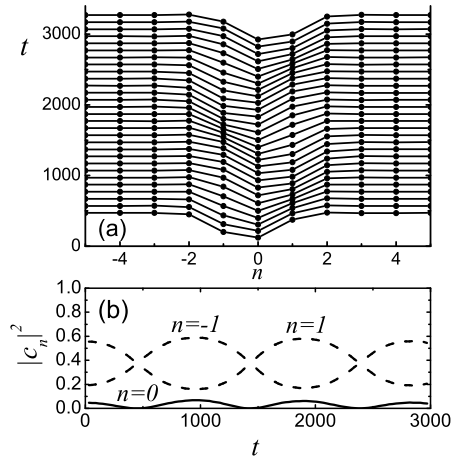}
\caption{Kink with oscillatory tail in Case 3 ($W_1=0$). (a)
Space-time evolution of $|c_n(t)|^2$ showing the dynamics of the
kink with oscillatory tail placed at $t=0$ asymmetrically with
respect to the lattice. The kink with oscillatory tail undergoes
periodic oscillations in the vicinity of the stable on-site
configuration. (b) Time variation of the norm of the central
particle and its two nearest neighbors. Parameters: $W_1=0$,
$W_2=0.0136$, $\omega=-0.88$. The kink with oscillatory tail has
frequency close to $\omega_f =-0.8896$ and it becomes unstable far
from this line (see Fig. \ref{Figure15}).} \label{Figure9}
\end{figure}

\subsection{ Soliton solutions } \label{Sec:solitons}

case of
It is well-known \cite{spa,ABK} that the standard DNLS equation (\ref{case5})
possesses multiple branches of localized solutions  which are
parametrized by the frequency detuning outwards the phonon band. Multiple
branches of the stationary localized solutions were also obtained in
a different model in the
presence of both linear and nonlinear latices~\cite{BluKon}. This allows one
to conjecture that any of the lattices introduced in Sec.~\ref{model}
should also possess multiple branches of the localized solutions and
kinks (discrete dark solitons).
Although a thorough study of each of the cases is by itself a
nontrivial problem that requires lengthy considerations, to present a
panoramic view of the possible nonlinear modes in the above lattices, in the
present paper we restrict our study to
some representative examples, referring, in most cases, to the lowest
branches.

{\em Case 2 ($W_2=0$).}

For the nonlinearity of the Case 2, using the method described in Sec. \ref{Sec:tail}, we obtain examples of localized
solutions of the Eq. (\ref{case23}), namely: pulse, staggered pulse, kink, staggered kink, wave (an anti-dark structure in the form of an oscillatory wave
on a non-zero background), and staggered wave. The
non-staggered solutions are presented in Fig. \ref{Figure2} for such a parameters: (a) $W_1=-0.0112$, $\omega=-0.99$,
(b) $W_1=0.0112$, $\omega=-0.9$, (c) $W_1=0.0112$, $\omega=-0.75$. The corresponding staggered solutions can be
constructed using the staggering transformation, described above. On-site equilibrium structures are shown but one can
also obtain the inter-site ones.

In Fig. \ref{Figure3}, the regions of existence of various solutions are indicated on the parameter plane
$(W_1,\omega)$. Solutions presented in Fig. \ref{Figure2} (a) to (c) have parameters shown by dots marked by the
capital letters A to C, respectively. Recall that pulses can exist in the two frequency ranges, $\omega<
\omega_b$ and $\omega> \omega_a$, for any $W_1$. However, they were found only for $W_1<0$, while for $W_1>0$ the
iterations initiated by the tail solution resulted in unbounded structures. In the portion of the plane with $W_1>0$
and $\omega_b < \omega < \omega_c$ kinks [see Fig. \ref{Figure2} (b)] and staggered waves were obtained. On the other
hand, in the portion with $W_1>0$ and $\omega_d < \omega < \omega_a$, we could construct staggered kinks and waves [see
panel (c) of Fig. \ref{Figure2}]. It is important to note that $|z|$ is close to 1 near the lines $\omega= \omega_a$
and $\omega= \omega_b$, where pulses, kinks, and waves were found to be wide; $z$ diverges (or vanishes) at $\omega=
\omega_c$ for staggered carrier field and at $\omega= \omega_d$ for non-staggered carrier field; in the range of
$\omega_d <\omega< \omega_c$, $z$ is always negative and close to 0, resulting in rapidly oscillating or sharply
localized (and typically unstable) solutions found from the tail construction.

Case 3 ($W_1=0$).

In the Case 3, three examples of the non-staggered stationary soliton solutions presented in Fig. \ref{Figure4} were
found. Shown are: (a) pulse, (b) kink, and (c) kink with oscillatory tail, referred to as oscillatory kink. On-site
structures are presented here, but inter-site ones can also be constructed. Model parameters corresponding to panels
(a) to (c) of Fig. \ref{Figure4} are depicted by dots in Fig. \ref{Figure5} marked by the capital letters A to C,
respectively. Parameters for the non-staggered solutions are: (a) $W_2=-0.0136$, $\omega=-0.9704$, (b) $W_2=0.0136$,
$\omega=-0.9$, (c) $W_2=0.0136$, $\omega=-0.85$, while the corresponding staggered solutions, as in the Case 2, can be
obtained using the staggering transformation.

The pulse tail solution (with $z>0$) exists for $\omega < \omega_b$
but the pulse itself exists in this region only for $W_2<0$, while
for positive $W_2$ we obtained unbounded solutions. Similarly,
the staggered pulse tail solution (with $z<0$) exists for $\omega >
\omega_a$ but the staggered pulse itself exists in this region
only for $W_2>0$, while negative $W_2$ leads to unbounded
solutions.

The kink exists for $W_2>0$ and $\omega_b < \omega < \omega_f$,
while the staggered kink for $W_2<0$ and $\omega_e < \omega <
\omega_a$. Finally, the oscillatory kink exists for $W_2>0$ and
$\omega_f < \omega < \omega_0$, while oscillatory staggered kink
for $W_2<0$ and $\omega_0 < \omega < \omega_e$.

In Fig. \ref{Figure6} and Fig. \ref{Figure7} we contrast the
behavior of pules in Cases 2 and 3, respectively. In both figures
(a) shows the space-time evolution of $|c_n(t)|^2$, while (b)
shows time variation of the norm of the central particles. On
purpose, we did not optimize the choice of $C_{\pm}$ to get
equilibrium on-site or inter-site solutions. As a result, the ensuing
profiles are non-stationary and, due to the presence of the
Peierls-Nabarro potential, they oscillate in the vicinity of
stable configurations. It turns out that in the Case 2 (Case 3)
the inter-site (on-site) configuration is stable. This conclusion
will be confirmed in Sec. \ref{Sec:stability}.
This illustrates the interesting phenomenon of potential inversion of
stability (cf. \cite{OJE}) in comparison with the standard DNLS mode
\cite{DNLS_review}.
Parameters in Fig. \ref{Figure6} are:
$W_1=-0.012$, $W_2=0$, $\omega= -0.99$, which
corresponds to the point A in Fig. \ref{Figure3}. Parameters in
Fig. \ref{Figure7} are: $W_1=0$, $W_2=-0.0136$, $\omega= -0.9704$,
which corresponds to the point A in Fig. \ref{Figure5}.

Similar results for dark solitons are presented in Fig.
\ref{Figure8} (kink in Case 2) and Fig. \ref{Figure9} (kink with
oscillatory tail in Case 3). One can see that in both cases the
on-site structures are stable and this will be confirmed in Sec.
\ref{Sec:stability}. Parameters in Fig. \ref{Figure8} are:
$W_1=0.012$, $W_2=0$, $\omega=-0.9$, which corresponds to the
point C in Fig. \ref{Figure3}. Parameters in Fig. \ref{Figure9}
are: $W_1=0$, $W_2=0.0136$, $\omega=-0.88$.

\begin{figure}
\includegraphics{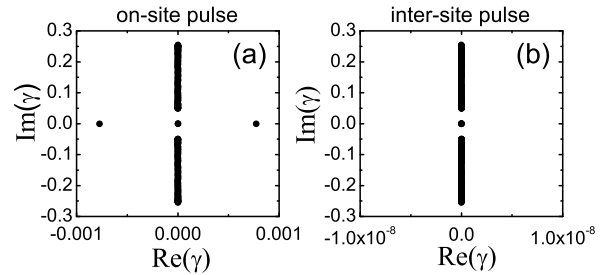}\caption{Spectra of (a) unstable on-site
and (b) stable inter-site pulses. The parameters (Case 2):
$W_1=-0.012$, $W_2=0$, $\omega= -0.99$ correspond to the
point A in Fig. \ref{Figure3}.} \label{Figure10}
\end{figure}

\begin{figure}
\includegraphics{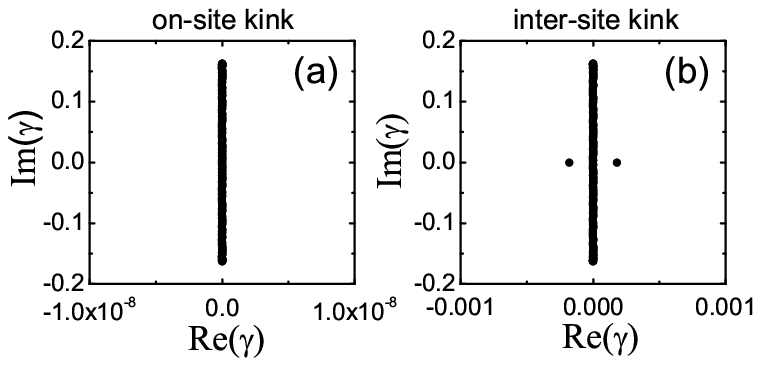}\caption{Spectra of (a) stable on-site
and (b) unstable inter-site kinks. The parameters (Case 2):
$W_1=0.012$, $W_2=0$, $\omega= -0.9$ correspond to the
point B in Fig. \ref{Figure3}.} \label{Figure11}
\end{figure}

\begin{figure}
\includegraphics{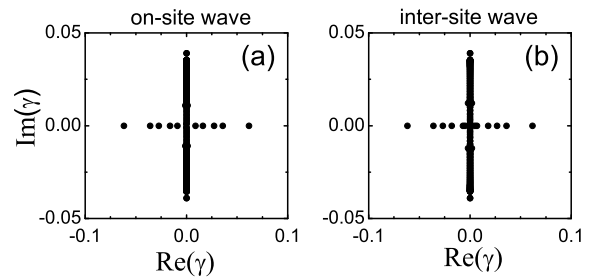}\caption{Spectra of (a) unstable on-site
and (b) unstable inter-site wave. The parameters (Case 2):
$W_1=0.012$, $W_2=0$, $\omega= -0.75$ correspond to the
point C in Fig. \ref{Figure3}.} \label{Figure12}
\end{figure}

\subsection{ Stability of soliton solutions }
\label{Sec:stability}

Let us now study the stability of the stationary soliton solutions
of the form of Eq. (\ref{StationarySolution}) described in Sec.
\ref{Sec:solitons}. We consider the following perturbed form of
the solutions,
\begin{eqnarray}
   c_n(t)=\left[f_n+\epsilon_n(t)\right] {\rm e}^{-i\omega t},
   \label{Perturbation}
\end{eqnarray}
where the small complex perturbation $\epsilon_n(t)$ is expressed
as follows,
\begin{eqnarray}
   \epsilon_n(t)=a_n(t)+ib_n(t).
   \label{EpsPerturbation}
\end{eqnarray}
Substituting Eq. (\ref{Perturbation}) into Eq.
(\ref{eq:psi1d-exp1}) we find that $\epsilon_n(t)$ is governed by
the following linearized equation,
\begin{eqnarray} \label{Linearized}
  i\dot \epsilon _n  = (\omega _0  - \omega)\epsilon _n
  + \omega _1 (\epsilon _{n - 1}  + \epsilon _{n + 1}) \nonumber \\
  + W_1 \Big[f_{n - 1}^2 Z_{n - 1}+ 2f_n (f_{n - 1}+f_{n +1})Z_n  \nonumber \\
  + f_{n + 1}^2 Z_{n + 1}  + f_n^2 (Z_{n - 1}+ Z_{n + 1}) \Big] \nonumber \\
  + W_2 \Big[ 2f_n f_{n - 1} Z_{n - 1}  + (f_{n - 1}^2  + f_{n + 1}^2)Z_n \nonumber \\
  + 2f_n f_{n + 1} Z_{n + 1}  \Big],
\end{eqnarray}
where $Z_n=3a_n+ib_n$. Then, separating real and imaginary parts
of Eq. (\ref{Linearized}) we derive the following system,
\begin{eqnarray}
  \left( {\begin{array}{*{20}c}
    {{\mathbf{\dot b}}}  \\
    {{\mathbf{\dot a}}}  \\
  \end{array} } \right)
  = \left( {\begin{array}{*{20}c}
    0 & \mathbf{K}  \\
    \mathbf{J} & 0  \\
  \end{array} } \right)
  \left( {\begin{array}{*{20}c}
    {\mathbf{b}}  \\
    {\mathbf{a}}  \\
  \end{array} } \right),
  \label{EigProblem}
\end{eqnarray}
where vectors $\mathbf{a}$ and $\mathbf{b}$ contain $a_n$ and
$b_n$, respectively, while the nonzero coefficients of matrices
$\mathbf{K}$ and $\mathbf{J}$ are given by,
\begin{eqnarray} \label{KnnJnn}
 K_{n,n - 1} &=& - \omega _1  - 3W_1 (f_{n - 1}^2+f_n^2)
 - 6W_2 f_{n - 1} f_n , \nonumber\\
 K_{n,n} &=& \omega - \omega_0 - 6W_1 f_n (f_{n - 1}+f_{n + 1})
 \nonumber \\ &-& 3W_2 (f_{n - 1}^2  + f_{n + 1}^2),  \nonumber\\
 K_{n,n + 1} &=& - \omega _1  - 3W_1 (f_n^2  + f_{n + 1}^2)
 - 6W_2 f_n f_{n + 1} ,  \nonumber\\
 J_{n,n - 1} &=& \omega _1  + W_1 (f_{n - 1}^2  + f_n^2)
 + 2W_2 f_{n - 1} f_n ,  \nonumber\\
 J_{n,n} &=& \omega_0 - \omega + 2W_1 f_n (f_{n - 1} + f_{n + 1})
 \nonumber \\ &+& W_2 (f_{n - 1}^2  + f_{n + 1}^2),  \nonumber\\
 J_{n,n + 1} &=& \omega _1  + W_1 (f_n^2  + f_{n + 1}^2)
 + 2W_2 f_n f_{n + 1} .
\end{eqnarray}
In the above expressions, $n=1,...,{\cal{N}}$, where ${\cal{N}}$
is the number of lattice points. For pulses and kinks, we used
periodic and anti-periodic boundary conditions, respectively.

A stationary solution is characterized as linearly stable if and
only if the eigenvalue problem
\begin{eqnarray}
   \left( {\begin{array}{*{20}c}
   0 & {\mathbf{K }}  \\    {\mathbf{J}} & 0  \\
   \end{array} } \right) \left( {\begin{array}{*{20}c} {\mathbf{b}}  \\
{\mathbf{a}} \\ \end{array} } \right)= \gamma \left( {\begin{array}{*{20}c} {\mathbf{b}}  \\
{\mathbf{a}} \\ \end{array} } \right)
   \label{EigValProblem}
\end{eqnarray}
results in nonpositive real parts of all eigenvalues $\gamma$.

\begin{figure}
\includegraphics{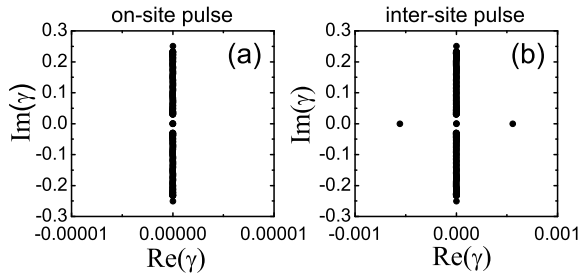}\caption{Spectra of (a) stable on-site
and (b) unstable inter-site pulses. Oscillatory motion of the
pulse in the vicinity of the on-site configuration is shown in
Fig. \ref{Figure6}. The parameters (Case 3) $W_1=0$, $W_2=-0.0136$,
$\omega= -0.9704$ correspond to the point A in Fig.
\ref{Figure5}.} \label{Figure13}
\end{figure}

\begin{figure}
\includegraphics{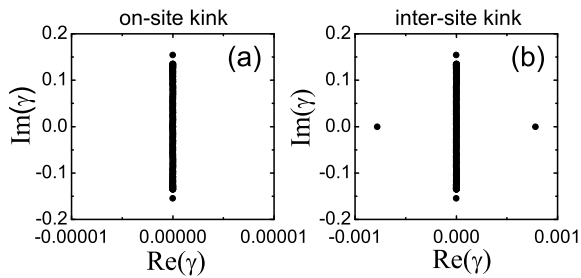}\caption{Spectra of (a) stable on-site
and (b) unstable inter-site kinks. The parameter values (Case 3)
$W_1=0$, $W_2=0.0136$, $\omega= -0.9$ correspond to the point B in
Fig. \ref{Figure5}.} \label{Figure14}
\end{figure}

\begin{figure}
\includegraphics{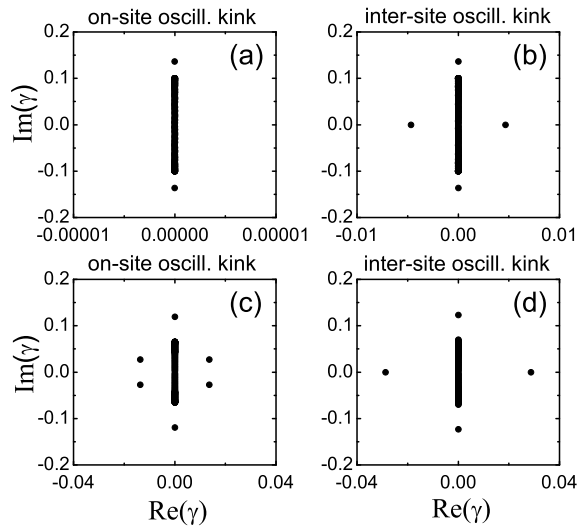}\caption{The top panels show spectra
of (a) stable on-site and (b) unstable inter-site oscillatory
kinks with $\omega= -0.88$. The bottom panels show the same for
$\omega= -0.85$ when both on-site and inter-site oscillatory kinks
become unstable. Note that the type of instability in (c) is
different from that in (d). In (c) there are two pairs of
complex-conjugate eigenvalues, while in (d) there is a pair of
purely real eigenvalues. The rest of the parameters are (Case 3)
$W_1=0$, $W_2=0.0136$.} \label{Figure15}
\end{figure}

The results of the stability analysis for the equilibrium
structures reported in Sec. \ref{Sec:solitons} are presented in
Figs. \ref{Figure10} to \ref{Figure12} for the Case 2 and in Figs.
\ref{Figure13} to \ref{Figure15} for the Case 3. The presented
spectra contain (i) the vibration frequencies of the homogeneous
background given by Eq. (\ref{q0}); (ii) a pair of zero-frequency
modes corresponding to the invariance with respect to the phase
shift; (iii) they also can include soliton internal modes falling
outside the phonon band, see, e.g., Fig. \ref{Figure14} (a). As
was already mentioned, the spectra of stable structures do not
possess eigenvalues with positive real parts, while those of the
unstable ones have such eigenvalues. Now we turn to the discussion
and comparison of the spectra in the Cases 2 and 3.

Case 2 ($W_2=0$).

Spectra of the on-site and inter-site pulses are presented in Fig. \ref{Figure10} (a) and (b), respectively. Interestingly, the inter-site
configuration is stable while the on-site one is unstable. This type of instability is typical for the discrete system with Peierls-Nabarro
potential, although in the standard cubic onsite nonlinearity case, the  results are entirely reversed in comparison to the present case \cite{DNLS_review}
(e.g., the
on-site pulse is stable, while the inter-site features a real eigenvalue
pair). This indicates that the Case 2 nonlinearity results in the shape of
the Peierls-Nabarro potential having a maximum (minimum) for the on-site (inter-site) pulses. Notice that as discussed in \cite{OJE}, such inversions of
stability may occur in such generalized models, upon varying their relevant
parameters (such as $W_1$ in the present case).
We will see that for the Case 3 nonlinearity the
situation for the pulse is exactly the opposite. The parameters used in this case are $W_1=-0.012$, $W_2=0$, $\omega= -0.97$, corresponding to the
point A in Fig. \ref{Figure3}.

Figure \ref{Figure11} shows results for the kink structures: the
on-site kink in (a) is stable while the inter-site one in (b) is
unstable. Here the location of maxima and minima of the
Peierls-Nabarro potential is the same as in the classical
discretization. The parameters $W_1=0.012$, $W_2=0.0$, $\omega= -0.9$
correspond to the point B in Fig. \ref{Figure3}.

Finally, in Fig. \ref{Figure12} we show that (a) the on-site wave and (b) the inter-site wave are both unstable. The
parameters $W_1=0.012$, $W_2=0$, and $\omega= -0.75$ correspond to the point C in Fig. \ref{Figure3}.

Case 3 ($W_1=0$).

Results of the stability analysis are presented in Figs. \ref{Figure13}-\ref{Figure15} for the three soliton solutions
displayed in panels (a) to (c) of Fig. \ref{Figure4}, respectively. The left panels show the spectra of the on-site
structures and the right panels show the same for the corresponding inter-site structures. One can see that, in
contrast to the Case 2, where the inter-site pulse was found to be stable, in the Case 3 the inter-site structures are
always unstable (this is analogous to the case of the standard cubic discrete model with the on-site nonlinearity).
This indicates that in the Case 3, the on-site (inter-site) structures are situated in the wells (on the peaks) of the
Peierls-Nabarro potential. On the other hand, panels (a) in Fig. \ref{Figure13} to Fig. \ref{Figure15} present purely
imaginary spectra for the on-site configurations, and this indicates that, for the corresponding values of model
parameters, all three types of equilibrium solutions are stable.
However, in Fig. \ref{Figure15} we demonstrate that the on-site
configuration of the oscillatory kink is stable at $\omega= 0.88$
[see panel (a)] but it can become unstable e.g. at $\omega= 0.85$ [see
panel (c)] with other parameters being unchanged and equal to
$W_1=0$, $W_2=0.0136$.

\section{ Conclusions }  \label{Sec:conclusions}

In this work, we have illustrated the potential for formulation of
a rich variety of tight-binding nonlinear lattice dynamical
models, stemming from the complex interplay of linear and
nonlinear periodic potentials in the Gross-Pitaevskii equation. We
have examined various particular possibilities, including
quasi-linear models, as well as strongly nonlinear models where
the nature of the coupling between the neighbors is itself
nonlinear. Furthermore, in the nonlinear ones among our models we
have studied the potential for the existence and stability, as
well as the dynamics of localized solutions. More specifically, we
have reported that the discrete model of Eq. (\ref{case23}) with
$W_1 \neq 0$, $W_2=0$ (Case 2), and $W_1=0$ and $W_2\neq 0$ (Case
3) supports a number of localized stationary solutions depicted in
Fig. \ref{Figure2} and Fig. \ref{Figure4}. Interestingly, the
nonlinearity of the Case 2 results in stable {\em inter-site}
pulse and stable {\em on-site} kink (both in staggered and
non-staggered forms). On the other hand, the nonlinearity of the
Case 3 supports only {\em on-site} stable localized solutions of
three different types, namely, pulses, kinks, and kinks with
oscillatory tails (all in both staggered and non-staggered forms).
These results suggest that in the discrete models with nonlinear
terms including interactions between nearest neighbors the profile
of the Peierls-Nabarro potential can change qualitatively
depending on the structure of the nonlinear terms and on the type
of the coherent structure. In fact, it has been demonstrated that
there exists a wide class of non-integrable discrete models of
this sort where the Peierls-Nabarro potential is precisely equal
to zero and equilibrium stationary solutions can be placed
anywhere with respect to the lattice points \cite{TImodels}.

A very natural extension of the present work would be to consider
similar types of reductions in higher dimensional settings and
to examine in detail the particular localized solutions that may
emerge in the resulting lattice models. In particular, higher
dimensionality may offer the potential for solutions with topological
charge; it would therefore be relevant to examine under what conditions
such solutions may be stable and how the relevant results relate to
the original continuum model.

\acknowledgments

YVB was supported by the FCT grant SFRH/PD/20292/2004. The work PGK
is supported by NSF-DMS-0505663 and
NSF-DMS-0619492, and NSF-CAREER and by the
University of Massachusetts. VVK acknowledges support from Ministerio de
Educaci\'on y Ciencia (MEC, Spain) under the
grant SAB2005-0195. The work of YVB and VVK was supported
by the FCT and European program FEDER under the grant
POCI/FIS/56237/2004.

\end{document}